\documentclass[twocolumn,showpacs,preprintnumbers,amsmath,amssymb,prb]{revtex4}

\usepackage{graphicx}
\usepackage{dcolumn}
\usepackage{bm}
\usepackage{color}

\begin{document}

\preprint{}

\title{Orbital- and spin-driven lattice instabilities in quasi-one-dimensional CaV$_2$O$_4$}

\author{T. Watanabe$^1$}
\thanks{Electronic address: tadataka@phys.cst.nihon-u.ac.jp}
\author{S. Kobayashi$^1$}
\author{Y. Hara$^2$}
\author{J. Xu$^3$}
\author{B. Lake$^3$}
\author{J.-Q. Yan$^4$}
\thanks{Present address: Materials Science and Technology Division, Oak Ridge National Laboratory, Oak Ridge, Tennessee 37831, USA}
\author{A. Niazi$^4$}
\thanks{Present address: Department of Physics, Jamia Millia Islamia, New Delhi 110025, India}
\author{D. C. Johnston$^{4,5}$}
\affiliation{$^1$Department of Physics, College of Science and Technology (CST), Nihon University, Chiyoda, Tokyo 101-8308, Japan}
\affiliation{$^2$National Institute of Technology, Ibaraki College, Hitachinaka 312-8508, Japan}
\affiliation{$^3$Helmholtz-Zentrum Berlin f\"ur Materialien und Energie, Hahn-Meitner-Platz 1, D-14109 Berlin, Germany}
\affiliation{$^4$Ames Laboratory, Iowa State University, Ames, Iowa 50011, USA}
\affiliation{$^5$Department of Physics and Astronomy, Iowa State University, Ames, Iowa 50011, USA}
\date{\today}

\begin{abstract}
Calcium vanadate CaV$_2$O$_4$ has a crystal structure of quasi-one-dimensional zigzag chains composed of orbital-active V$^{3+}$ ions and undergoes successive structural and antiferromagnetic phase transitions at $T_s\sim 140$ K and $T_N \sim 70$ K, respectively. We perform ultrasound velocity measurements on a single crystal of CaV$_2$O$_4$. The temperature dependence of its shear elastic moduli exhibits huge Curie-type softening upon cooling that emerges above and below $T_s$ depending on the elastic mode. The softening above $T_s$ suggests the presence of either onsite Jahn-Teller-type or intersite ferro-type orbital fluctuations in the two inequivalent V$^{3+}$ zigzag chains. The softening below $T_s$ suggests the occurrence of a dimensional spin-state crossover, from quasi-one to three, that is driven by the spin-lattice coupling along the inter-zigzag-chain orthogonal direction. The successive emergence of the orbital- and spin-driven lattice instabilities above and below $T_s$, respectively, is unique to the orbital-spin zigzag chain system of CaV$_2$O$_4$.
\end{abstract}


\maketitle

\section{Introduction}

Orbitally degenerate frustrated magnets have attracted considerable interest because they display a variety of complex ground states with unusual magnetic and orbital orders [\cite{Radaelli}]. The prototypical examples are vanadate spinels $A$V$_2$O$_4$ with divalent $A^{2+}$ ions such as Zn$^{2+}$, Mg$^{2+}$, and Cd$^{2+}$, where the trivalent magnetic V$^{3+}$ ions are characterized by double occupancy of the triply degenerate $t_{2g}$ orbitals, and form a pyrochlore lattice. With cooling, the vanadate spinels undergo successive structural and antiferromagnetic (AF) phase transitions at a temperature $T_s$ and a lower temperature $T_N<T_s$ [\cite{Lee,Mamiya,Giovannetti}]. For $A$V$_2$O$_4$, the structural phase transition is understood to arise from a long-range ordering of the V $t_{2g}$ orbitals in which the lowering of the lattice symmetry results in the release of frustration (magnetic ordering).

\begin{figure}[b]
\begin{center}
\includegraphics[scale=0.4]{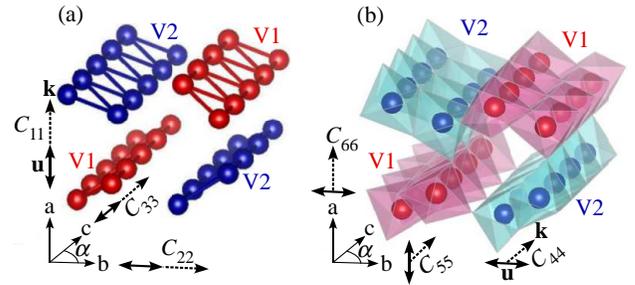}
\caption{\label{fig:fig1} (Color online) Bonding network in the crystal structure of CaV$_2$O$_4$: (a) V atoms and (b) VO$_6$ octahedra. In (a) and (b), V1 and V2 denote the inequivalent V sites. The angle $\alpha$ displayed in the crystal axes corresponds to the monoclinic angle in the monoclinic crystal phase below $T_s$ [\cite{Niazi}]. In (a), the propagation vector {\bf k} and polarization vector {\bf u} of the longitudinal sound waves for $C_{11}$, $C_{22}$, and $C_{33}$ are indicated. In (b), the propagation vector {\bf k} and polarization vector {\bf u} of the transverse sound waves for $C_{44}$, $C_{55}$, and $C_{66}$ are indicated.}
\end{center}
\end{figure}

\begin{figure}[t]
\begin{center}
\includegraphics[scale=0.45]{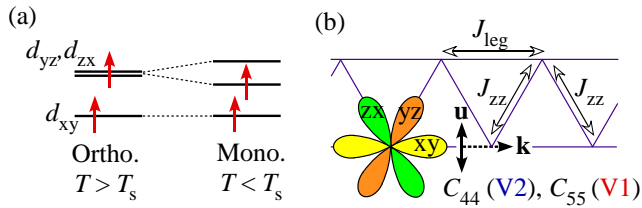}
\caption{\label{fig:fig2} (Color online) (a) Schematic $t_{2g}$ energy levels of V$^{3+}$ 3$d^2$ electrons in the orthorhombic (left) and monoclinic (right) crystal phases of CaV$_2$O$_4$. (b) Schematic of the V $t_{2g}$ orbitals in the zigzag chain of CaV$_2$O$_4$. In (b), two competing AF interactions of the nearest-neighbor $J_{zz}$ along the zigzags and the next-nearest-neighbor $J_{leg}$ along the leg are indicated. Additionally, the propagation vector {\bf k} and polarization vector {\bf u} of the transverse sound wave corresponding to the $\epsilon_{xx}-\epsilon_{yy}$ elastic mode ($d_{yz}/d_{zx}$-active mode) for the $t_{2g}$ orbitals are indicated. For the V1 and V2 zigzag chains of CaV$_2$O$_4$, $C_{55}$ and $C_{44}$, respectively, correspond to this elastic mode.}
\end{center}
\end{figure}

Unlike the vanadate spinels $A$V$_2$O$_4$, calcium vanadate CaV$_2$O$_4$ crystallizes in the orthorhombic CaFe$_2$O$_4$-type structure (space group $Pnam$) at room temperature, which consists of the two inequivalent V$^{3+}$ zigzag chains running along the crystal $c$ axis [Fig. 1(a)]. Let V1 and V2 denote the inequivalent V sites. The magnetic V$^{3+}$ ions on the V1 and V2 sites possess spin $S=1$ because of two unpaired 3$d$ electrons, where magnetism is dominated by AF interactions [\cite{Zong,Niazi,Pieper}]. With the V1 and V2 zigzag chains consisting of V$^{3+}$ triangular loops, CaV$_2$O$_4$ is expected to be a quasi-one-dimensional frustrated magnet.

In CaV$_2$O$_4$, the V$^{3+}$ ions are surrounded by slightly distorted O octahedra, which share edges within the V1 and V2 zigzag chains but share corners between the V1 and V2 zigzag chains [Fig. 1(b)]. The trivalent magnetic V$^{3+}$ ion in the octahedral O environment is characterized by double occupancy of the triply-degenerate $t_{2g}$ orbitals. In CaV$_2$O$_4$ with the high-temperature orthorhombic crystal structure, the octahedral O environment is tetragonally compressed, and the $t_{2g}$ orbitals are split into a lower nondegenerate $d_{xy}$ orbital and higher doubly-degenerate $d_{yz}$ and $d_{zx}$ orbitals [Fig. 2(a)] [\cite{Pieper,Distortion}]. Therefore, although the O octahedra in the orthorhombic CaV$_2$O$_4$ is slightly distorted, the V$^{3+}$ ions should have an orbital degree of freedom [Fig. 2(a)]; the lower $d_{xy}$ orbital is occupied by one of the two electrons, and the higher doubly-degenerate $d_{yz}$ and $d_{zx}$ orbitals are partially occupied by the remaining one electron. Fig. 2(b) depicts a schematic of the V $t_{2g}$ orbitals in the zigzag chain of CaV$_2$O$_4$. The nondegenerate $d_{xy}$ orbital aligns with the leg direction of the zigzag chain, but the doubly-degenerate $d_{yz}$ and $d_{zx}$ orbitals align to the zigzag directions. In CaV$_2$O$_4$, its magnetism is understood to be dominated by two competing AF interactions of the nearest-neighbor $J_{zz}$ along the zigzags and the next-nearest-neighbor $J_{leg}$ along the leg [Fig. 2(b)]. Hence, the ground state of CaV$_2$O$_4$ has attracted interest as a quasi-one-dimensional orbitally degenerate frustrated magnet [\cite{Zong,Niazi,Pieper,Kikuchi,Chern,Chern2,Nersesyan,Sela,Pchelkina}].

CaV$_2$O$_4$ undergoes successive structural and magnetic phase transitions; a weak orthorhombic-to-monoclinic lattice distortion at a temperature $T_s\sim 140$ K, and an AF ordering at a lower temperature $T_N \sim 70$ K, $T_s>T_N$ [\cite{Zong,Niazi,Pieper}]. In the monoclinic crystal structure below $T_s$, the monoclinic angle is between the $b$ and $c$ axes ($\alpha$) of the orthorhombic crystal structure above $T_s$ [Figs. 1(a) and 1(b)], which evolves continuously below $T_s$ and saturates at $\alpha\simeq$ 89.2$^{\circ}$ at low temperatures [\cite{Niazi,Pieper,Niazi2}]. Moreover, the structural distortion below $T_s$ lifts the orbital degeneracy [Fig. 2(a)]. From the magnetic susceptibility and the neutron scattering measurements in the single crystal of CaV$_2$O$_4$, each zigzag spin chain of this compound [Fig. 2(b)] evidently changes state at $T_s$ from the $J_{leg}$-dominant Haldane-chain state above $T_s$ to the $J_{leg}/J_{zz}$-competing spin-ladder state below $T_s$, which results from the $J_{zz}$ enhancement driven by a ferro-type ordering of the $d_{yz}/d_{zx}$ orbitals below $T_s$ [\cite{Pieper}].

Below $T_N$, CaV$_2$O$_4$ exhibits a three-dimensional AF order with a propagation vector {\bf Q} = (0,$\frac{1}{2}$,$\frac{1}{2}$) despite the quasi-one-dimensional character of the crystal structure [\cite{Zong,Niazi,Pieper,Hastings,Bertaut,Sugiyama}]. Here, the AF structure consists of collinear V1/V2 zigzag chains that are canted with respect to each other [\cite{Zong,Pieper}], where the ordered magnetic moment of $1.0\mu_B\le\mu\le1.59\mu_B$ reduced from 2$\mu_B$ for $S$ = 1 is considered to arise from the low-dimensional and/or frustrated magnetic character [\cite{Zong,Niazi,Pieper,Hastings,Bertaut,Sugiyama}]. For CaV$_2$O$_4$, the orbital ordering at $T_s$ in the V$^{3+}$ zigzag chains is believed to lead to the three-dimensional AF ordering (due to the release of frustration) at the lower $T_N$.

In this paper, we present ultrasound velocity measurements of the quasi-one-dimensional orbital degenerate CaV$_2$O$_4$, from which we determine the elastic moduli of this compound. The elastic modulus of a crystal is a thermodynamic tensor quantity, and therefore the ultrasound velocity measurements of the symmetrically independent elastic moduli in a crystal can provide symmetry-resolved thermodynamic information. In magnets, the modified sound dispersions caused by magnetoelastic coupling allow one to extract detailed information about the interplay of the lattice, spin, and orbital degrees of freedom [\cite{Luthi,Kino,Kataoka,Hazama,Watanabe1,Nii,Watanabe2,Watanabe3,Watanabe4,Watanabe5,Watanabe6}]. For CaV$_2$O$_4$, we find two different types of orbital-driven elastic anomalies in the orthorhombic paramagnetic (PM) phase above $T_s$, and a spin-driven elastic anomaly in the lower-temperature monoclinic PM phase below $T_s$ that are observed in the symmetrically different elastic modes. These elastic anomalies should be precursors to successive occurrences of the orbital order at $T_s$ and the spin order at the lower $T_N$, a feature which is unique to the orbital-spin zigzag chain system of CaV$_2$O$_4$.

\section{Experimental}

\begin{table}[b]
\caption{\label{tab:table1} Elastic moduli for CaV$_2$O$_4$ with the orthorhombic crystal structure, and the corresponding sound mode (propagation vector {\bf k} and polarization vector {\bf u}) and symmetry.}
\begin{ruledtabular}
\begin{tabular}{ccc}
Elastic modulus & Sound mode ({\bf k} and {\bf u}) & Symmetry\\
\hline
$C_{11}$ & Longitudinal wave ({\bf k} $\parallel$ {\bf u} $\parallel$ {\bf a}) & A$_g$\\
$C_{22}$ & Longitudinal wave ({\bf k} $\parallel$ {\bf u} $\parallel$ {\bf b}) & A$_g$\\
$C_{33}$ & Longitudinal wave ({\bf k} $\parallel$ {\bf u} $\parallel$ {\bf c}) & A$_g$\\
\hline
$C_{44}$ & Transverse wave ({\bf k} $\parallel$ {\bf c}, {\bf u} $\parallel$ {\bf b}) & B$_{3g}$\\
$C_{55}$ & Transverse wave ({\bf k} $\parallel$ {\bf c}, {\bf u} $\parallel$ {\bf a}) & B$_{2g}$\\
$C_{66}$ & Transverse wave ({\bf k} $\parallel$ {\bf a}, {\bf u} $\parallel$ {\bf b}) & B$_{1g}$\\
\end{tabular}
\end{ruledtabular}
\end{table}

Single crystals of CaV$_2$O$_4$ with $T_s\sim$ 140 K and $T_N\sim$ 70 K were grown by the floating-zone method [\cite{Niazi}]. The ultrasound velocity measurements were performed using the phase-comparison technique with longitudinal and transverse sound waves at a frequency of 30 MHz. The ultrasound waves were generated and detected by LiNbO$_3$ transducers glued on the parallel mirror surfaces of the crystal which are respectively perpendicular to the $a$, $b$, and $c$ orthorhombic axes. Measurements were taken to determine the symmetrically independent elastic moduli in the orthorhombic crystal, specifically, $C_{11}$, $C_{22}$, $C_{33}$, $C_{44}$, $C_{55}$, and $C_{66}$ (see Table I). In Figs. 1(a) and 1(b), the propagation vector {\bf k} and polarization vector {\bf u} of the sound waves for the respective elastic moduli are indicated along with the bonding network of CaV$_2$O$_4$ crystal. As indicated in Fig. 1(a), the longitudinal sound wave corresponding to the compressive elastic modulus $C_{33}$ propagates along the V1 and V2 chains ({\bf k} $\parallel$ {\bf c}), whereas the longitudinal waves corresponding to the compressive $C_{11}$ and $C_{22}$ propagate orthogonal to the V1 and V2 chains ({\bf k} $\perp$ {\bf c}). Likewise, in Fig. 1(b), the transverse sound waves corresponding to the shear elastic moduli $C_{44}$ and $C_{55}$ propagate along the V1 and V2 chains ({\bf k} $\parallel$ {\bf c}), whereas the transverse wave corresponding to the shear $C_{66}$ propagates orthogonal to the V1 and V2 chains ({\bf k} $\perp$ {\bf c}). The sound velocities of CaV$_2$O$_4$ measured at room temperature (300 K) are 3510 m/s for $C_{11}$, 4200 m/s for $C_{22}$, 7740 m/s for $C_{33}$, 2970 m/s for $C_{44}$, 3620 m/s for $C_{55}$, and 4380 m/s for $C_{66}$.

\section{Results}

The temperature ($T$) dependence of the compressive elastic moduli $C_{11}(T)$, $C_{22}(T)$, and $C_{33}(T)$, respectively, in CaV$_2$O$_4$ all exhibit small discontinuous changes at $T_s$ and $T_N$ [marked by arrows in Figs. 3(a)-3(c)]. In regard to the magnetostructural phases of CaV$_2$O$_4$, specifically, the orthorhombic PM phase ($T>T_s$), monoclinic PM phase ($T_N<T<T_s$), and monoclinic AF phase ($T<T_N$), all three compressive elastic moduli exhibit monotonic hardening upon cooling, as is usually observed in solids [\cite{Varshni}].

\begin{figure}[t]
\begin{center}
\includegraphics[scale=0.45]{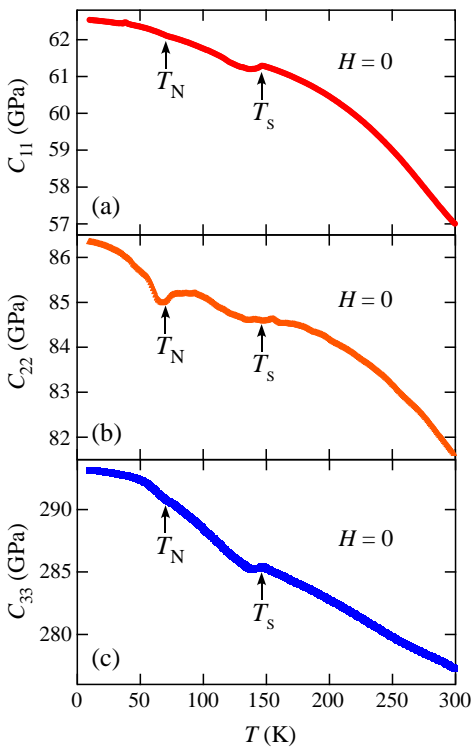}
\caption{\label{fig:fig3} (Color online) Compressive elastic moduli of CaV$_2$O$_4$ as functions of temperature. (a) $C_{11}(T)$, (b) $C_{22}(T)$, and (c) $C_{33}(T)$. The labelled arrows indicate $T_s$ and $T_N$.}
\end{center}
\end{figure}

\begin{figure}[t]
\begin{center}
\includegraphics[scale=0.45]{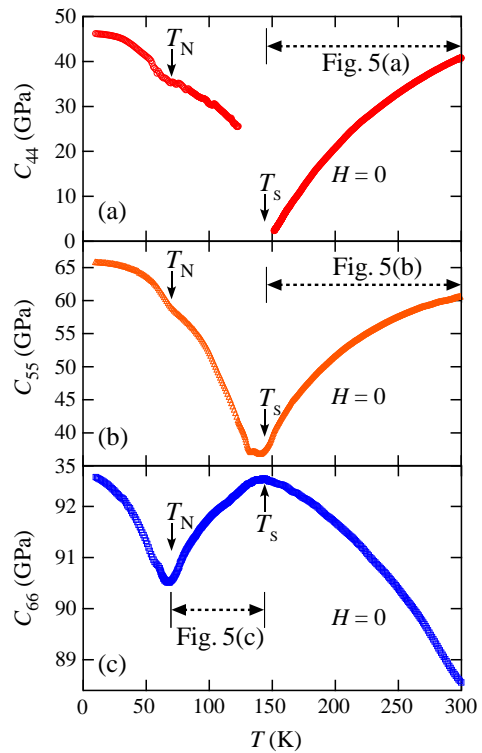}
\caption{\label{fig:fig4} (Color online) Shear elastic moduli of CaV$_2$O$_4$ as functions of temperature. (a) $C_{44}(T)$, (b) $C_{55}(T)$, and (c) $C_{66}(T)$. The labelled arrows indicate $T_s$ and $T_N$. The experimental data in the temperature ranges indicated by double-headed arrows are displayed rescaled in Fig. 5.}
\end{center}
\end{figure}

In contrast to the compressive elastic moduli (Fig. 3), the shear elastic moduli of CaV$_2$O$_4$ are found to exhibit unusual temperature variations that depend on the elastic mode. Figure 4 gives the temperature dependence of the shear elastic moduli $C_{44}(T)$, $C_{55}(T)$, and $C_{66}(T)$ in CaV$_2$O$_4$. In the orthorhombic PM phase ($T>T_s$), $C_{44}(T)$ [Fig. 4(a)] and $C_{55}(T)$ [Fig. 4(b)] exhibit huge Curie-type ($\sim -$1/$T$-type) softening upon cooling, whereas $C_{66}(T)$ [Fig. 4(c)] exhibits ordinary hardening. Conversely, in the monoclinic PM phase ($T_N<T<T_s$), $C_{66}(T)$ [Fig. 4(c)] exhibits Curie-type softening upon cooling, whereas $C_{44}(T)$ [Fig. 4(a)] and $C_{55}(T)$ [Fig. 4(b)] exhibit hardening upon cooling. Here, it is impossible to measure $C_{44}(T)$ below $T_s$ down to $\sim$125 K [Fig. 4(a)] because the ultrasound signal damped strongly. In the monoclinic AF phase ($T<T_N$), $C_{44}(T)$, $C_{55}(T)$, and $C_{66}(T)$ all exhibit hardening upon cooling (Fig. 4). For the shear elastic moduli of CaV$_2$O$_4$, $T_s$ and $T_N$ intriguingly correspond to the elastic-mode-dependent softening-hardening turning points (Fig. 4). In the next section, we shall discuss the origin of the elastic-mode-dependent unusual temperature variations of the shear elastic moduli.

In the present study, we measured $C_{11}(T)\sim C_{66}(T)$ not only in the absence of a magnetic field [Figs. 3 and 4], but also with magnetic fields up to 7 T. We find an absence of a magnetic field effect on any of the elastic properties.

\section{Discussion}

The present experimental results reveal that, in CaV$_2$O$_4$, while symmetry-conserving isotropic elastic modes of the compressive moduli $C_{11}(T)$, $C_{22}(T)$, and $C_{33}(T)$ exhibit ordinal hardening upon cooling (Fig. 3), symmetry-lowering anisotropic elastic modes of the shear moduli $C_{44}(T)$, $C_{55}(T)$, and $C_{66}(T)$ exhibit unusual elastic-mode-dependent temperature variations (Fig. 4). These elastic properties should reflect the low-dimensional orbital and spin characters of this compound, which strongly couple to the lattice. From here on, we shall discuss the origin of these unusual temperature variations in $C_{44}(T)$, $C_{55}(T)$, and $C_{66}(T)$.

First we address the origin of Curie-type softening in $C_{44}(T)$ and $C_{55}(T)$ in the orthorhombic PM phase ($T>T_s$) [Figs. 4(a) and 4(b)]. This elastic instability is quenched below $T_s$, and therefore should be a precursor to the structural transition at $T_s$ lifting the orbital degeneracy of V$^{3+}$ ions [Fig. 2(a)].

In an orbital-degenerate system, the temperature dependence of the elastic modulus $C_{\Gamma}(T)$ above the structural transition temperature is explained by assuming the coupling of the ultrasound to the orbital-degenerate ions through the onsite orbital-strain (quadrupole-strain) interaction, and the presence of the intersite orbital-orbital (quadrupole-quadrupole) interaction. [\cite{Luthi,Kino,Kataoka,Hazama}]. A mean-field expression of $C_{\Gamma}(T)$ for the orbital-degenerate system is given as
\begin{equation}
C_{\Gamma}(T) = C_{\Gamma}^0 \frac{T-T_c}{T-\theta},
\label{eq:Curie}
\end{equation}
with $C_{\Gamma}^0$ the background elastic constant, $T_c$ the second-order critical temperature for elastic softening $C_{\Gamma}\rightarrow$ 0, and $\theta$ the intersite orbital-orbital interaction. The difference of the two characteristic temperatures $T_c-\theta$ is the energy gain from the onsite orbital-strain interaction, which corresponds to the onsite Jahn-Teller coupling energy $E_{JT}$ with $T_c-\theta=E_{JT}$. $\theta$ is positive (negative) when the interaction is ferro-distortive (antiferro-distortive). The ferro-distortive (antiferro-distortive) interaction is expected to lead to a ferro-orbital order (antiferro-orbital order) with (without) a macroscopic lattice distortion.

Fits of the experimental $C_{44}(T)$ and $C_{55}(T)$ to Eq. (1) for $T>T_s$ [Figs. 5(a) and 5(b); solid black curves] reproduce very well the experimental data. Values for the fitting parameters are also presented. The larger $E_{JT}$ in $C_{44}(T)$ than in $C_{55}(T)$ indicates that the strain for $C_{44}$ generates a stronger onsite orbital-strain interaction than the strain for $C_{55}$. The change in sign of $\theta$ between $C_{44}(T)$ and $C_{55}(T)$ indicates the coexistence of different types of intersite orbital-orbital interactions. Moreover, the larger magnitude of $\theta$ in $C_{55}(T)$ than in $C_{44}(T)$ indicates that the intersite orbital-orbital interaction affecting $C_{55}(T)$ is stronger than that affecting $C_{44}(T)$.

\begin{figure}[b]
\begin{center}
\includegraphics[scale=0.45]{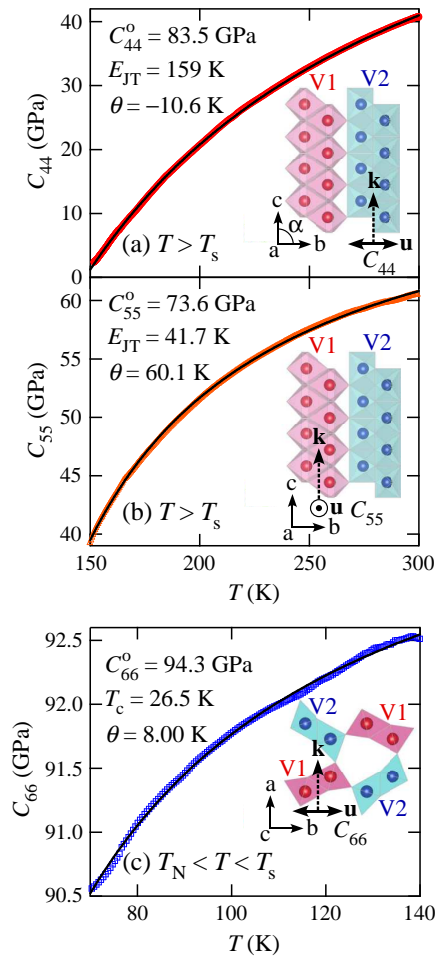}
\caption{\label{fig:fig5} (Color online) Expanded view of the shear elastic moduli for CaV$_2$O$_4$ given in from Fig. 4: (a) $C_{44}(T)$ in the orthorhombic PM phase ($T > T_s$), (b) $C_{55}(T)$ in the orthorhombic PM phase ($T > T_s$), and (c) $C_{66}(T)$ in the monoclinic PM phase ($T_N < T < T_s$). The solid black curves in all three plots are fits using Eq. (1) of the experimental data; values for the fitting parameters are also listed. The VO$_6$ octahedra of the V1 and V2 zigzag chains are illustrated along with the sound propagation vector {\bf k} and polarization vector {\bf u}. In (a), the angle $\alpha$ displayed in the crystal axes corresponds to the monoclinic angle in the monoclinic crystal phase ($T<T_s$) [\cite{Niazi}].}
\end{center}
\end{figure}

For CaV$_2$O$_4$, the monoclinic angle in the low-temperature monoclinic crystal structure is between the $b$ and $c$ axes ($\alpha$) of the high-temperature orthorhombic crystal structure [\cite{Niazi}]; see Fig. 5(a). Hence, with $C_{44}$ corresponding to the $\alpha$ tilt mode ({\bf k} $\parallel$ {\bf c} and {\bf u} $\parallel$ {\bf b}), Curie-type softening in $C_{44}(T)$ above $T_s$ should be a precursor to the orthorhombic-to-monoclinic lattice distortion at $T_s$. If $E_{JT}$ is much stronger than $\theta$ in $C_{44}(T)$ [Fig. 5(a)], the orthorhombic-to-monoclinic lattice distortion at $T_s$ in CaV$_2$O$_4$ should be a Jahn-Teller-type lattice distortion, where the onsite orbital-strain interaction lifts the orbital degeneracy.

Distinct from $C_{44}$, the strain generated by ultrasound in $C_{55}$ ({\bf k} $\parallel$ {\bf c} and {\bf u} $\parallel$ {\bf a}) tilts the angle between the $c$ and $a$ axes of the orthorhombic crystal structure [Fig. 5(b)]. Therefore, Curie-type softening in $C_{55}(T)$ above $T_s$ should be an elastic instability rather than the precursor to the $bc$-plane monoclinic lattice distortion at $T_s$. With the strong positive $\theta$ in $C_{55}(T)$ [Fig. 5(b)], Curie-type softening in $C_{55}(T)$ should be a precursor to an orbital ordering, where the intersite ferro-orbital interactions play an important role. At $T_s$, a small monoclinic lattice distortion within the $ac$-plane might coincide with a larger distortion within the $bc$-plane, which has not been experimentally identified so far.

We here note that, in CaV$_2$O$_4$, the inequivalent V1 and V2 sites each have different crystal-field $z$ directions of the 3$d$ orbitals, specifically, close to the crystal $b$ axis ($a$ axis) in the V1 (V2) sites [\cite{Pieper,Pchelkina}]. For the $t_{2g}$ orbitals of the $d$ electrons, the strain $\epsilon_{xx}-\epsilon_{yy}$ generated by transverse sound waves does not change the energy level of $d_{xy}$ orbital, but lowers/raises the energy levels of $d_{yz}$ and $d_{zx}$ orbitals, respectively. In CaV$_2$O$_4$, $C_{44}$ corresponds to such a $d_{yz}/d_{zx}$-active elastic mode for the V2 sites [Figs. 2(b) and 5(a)], and $C_{55}$ for the V1 sites [Figs. 2(b) and 5(b)]. Therefore, in CaV$_2$O$_4$, the strain generated by ultrasound should couple to the $d_{yz}$ and $d_{zx}$ orbitals of the V2 sites in regard to $C_{44}$ and the V1 sites in regard to $C_{55}$. Given the fitted parameter values of $E_{JT}$ and $\theta$ [Figs. 5(a) and 5(b)], Curie-type softening in $C_{44}(T)$ and $C_{55}(T)$ should be driven by two different types of orbital fluctuations: respectively, onsite Jahn-Teller-type orbital fluctuations in the V2 sites affecting $C_{44}(T)$ and intersite ferro-orbital fluctuations in the V1 sites affecting $C_{55}(T)$. Consequently, the observation of Curie-type softening in $C_{44}(T)$ and $C_{55}(T)$ above $T_s$ suggests that, at $T_s$, the Jahn-Teller-type orbital ordering occurs in the V2 sites, but another type of orbital ordering driven by the intersite ferro-orbital interactions occurs in the V1 sites. Here, the coincidental occurrence at $T_s$ of the two different types of orbital ordering at the respective V1 and V2 sites indicates that these orbital orderings are a cooperative phenomenon arising from the weak coupling of the two inequivalent V1 and V2 zigzag chains. Below $T_s$, the ferro-type orbital configurations should form in the inequivalent V1 and V2 zigzag chains [Fig. 6(a)]. These orbital configurations are compatible with the emergence of the spin-ladder state below $T_s$ in each zigzag spin chain, as indicated from the magnetic susceptibility and neutron scattering measurements [\cite{Pieper}].

Next we discuss the origin of Curie-type softening in $C_{66}(T)$ in the monoclinic PM phase ($T_N<T<T_s$) [Fig. 4(c)]. Note that, in the monoclinic PM phase ($T_N<T<T_s$), $C_{44}(T)$ and $C_{55}(T)$ exhibit ordinal hardening upon cooling [Figs. 4(a) and 4(b)], which indicates that the orbital degeneracies of the V1 and V2 sites are both lifted below $T_s$. Thus, taking into account that $C_{66}(T)$ exhibits ordinal hardening upon cooling in the orthorhombic PM phase ($T>T_s$) [Fig. 4(c)], the emergence of Curie-type softening in $C_{66}(T)$ in the monoclinic PM phase ($T_N<T<T_s$) should be the result of the generation of a new spin-lattice coupling in this phase, which is driven by the orbital ordering at $T_s$. Additionally, note that Curie-type softening in $C_{66}(T)$ is quenched below $T_N$ [Fig. 4(c)]. This indicates that the elastic instability of $C_{66}$ in the monoclinic PM phase ($T_N<T<T_s$) is a precursor to the AF transition at $T_N$.

\begin{figure}[b]
\begin{center}
\includegraphics[scale=0.45]{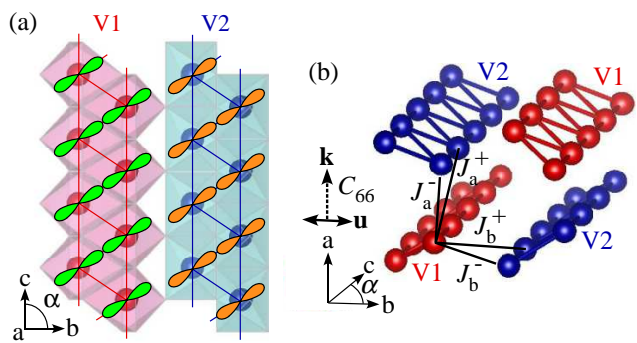}
\caption{\label{fig:fig6} (Color online) (a) Ferro-type orbital configuration of CaV$_2$O$_4$ projected onto the $bc$ plane, which displays VO$_6$ octahedra of the V1 and V2 zigzag chains. (b) V1 and V2 bonding networks indicating the inter-zigzag-chain neighboring exchange interactions $J_a^{\pm}$ and $J_b^{\pm}$. In (a) and (b), angle $\alpha$ shown with the crystal axes corresponds to that in the monoclinic crystal phase ($T<T_s$) [\cite{Niazi}]. In (b), the sound propagation vector {\bf k} and polarization vector {\bf u} for $C_{66}$ tilts the angle between the $a$ and $b$ axes.}
\end{center}
\end{figure}

In a spin-lattice coupled system, the temperature dependence of the elastic modulus $C_{\Gamma}(T)$ is explained by assuming a coupling of ultrasound with the magnetic ions through the magnetoelastic coupling acting on the exchange interactions, where the exchange striction arises from an ultrasound modulation of the exchange interactions [\cite{Luthi,Watanabe5,Watanabe6}]. In analogy to the orbital-degenerate system, Curie-type softening in $C_{\Gamma}(T)$ in the spin-lattice coupled system is explained by assuming the coupling of ultrasound to the magnetic ions via the exchange striction mechanism, and the presence of exchange-striction-sensitive intersite spin-spin interactions. The mean-field expression of the soft mode $C_{\Gamma}(T)$ for the spin-lattice coupled system should have the same form as in Eq. (1) with $C_{\Gamma}^0$ the background elastic constant, $T_c$ the second-order critical temperature for elastic softening $C_{\Gamma}\rightarrow$ 0, and $\theta$ the exchange-striction-sensitive intersite spin-spin interaction [\cite{Luthi,Watanabe5,Watanabe6}].

In Fig. 5(c), a fit of the experimental $C_{66}(T)$ to Eq. (1) in $T_N<T<T_s$ is drawn as a solid black curve, which reproduces very well the experimental data. Values for the fit parameters are also presented. The fitted parameter value of $T_c$ is lower than the experimentally observed N$\acute{e}$el temperature $T_N$, $T_c<T_N$, indicating that the phase transition at $T_N$ is of first order. Taking into account that the second-order transition at $T_N$ was suggested from the thermal expansion measurements [\cite{Niazi}], the transition at $T_N$ might be of weak first order. The positive fitted value of $\theta$ indicates the dominance of ferro-distortive intersite spin-spin interactions.

In CaV$_2$O$_4$, the ultrasound for $C_{66}$ generates the strain within the $ab$ plane, which is orthogonal to both the V1 and V2 zigzag chains [Fig. 5(c)]. Therefore the emergence of the precursor softening to the AF transition in $C_{66}(T)$ indicates that the three-dimensional AF ordering is driven in the quasi-one-dimensional spin system via the spin-lattice coupling along the inter-zigzag-chain orthogonal direction. That is, the generation of this "orthogonal-type" spin-lattice coupling in the monoclinic PM phase ($T_N<T<T_s$) should give rise to a spin-state crossover from quasi-one-dimension to three-dimensions, a feature which is unique to the orbital-spin zigzag chain system of CaV$_2$O$_4$.

For the three-dimensional AF ordering in CaV$_2$O$_4$, the inter-zigzag-chain exchange interactions should play a crucial role via the "orthogonal-type" spin-lattice coupling. As evident in Fig. 6(b), there are two types of inter-zigzag-chain neighboring exchange interactions in CaV$_2$O$_4$: $a$-axis stacking $J_a^{\pm}$ and $b$-axis stacking $J_b^{\pm}$. Curie-type softening in $C_{66}(T)$ indicates that $J_a^{\pm}$ and $J_b^{\pm}$ in the monoclinic PM phase ($T_N<T<T_s$) are sensitive to the monoclinic lattice deformation within the $ab$ plane [Fig. 6(b)]. Note here that the tilting of the monoclinic angle $\alpha$ below $T_s$ makes the $b$-axis stacking $J_b^+$ and $J_b^-$ become inequivalent, although the magnitude of the $\alpha$ tilting is very small [\cite{Niazi}]. Thus, strictly speaking, Curie-type softening in $C_{66}(T)$ should be driven by $J_a^{\pm}$ and the slightly inequivalent $J_b^+$ and $J_b^-$. For CaV$_2$O$_4$, it is expected that a small monoclinic lattice distortion within the $ab$ plane coincides with AF ordering, although the additional lattice distortion below $T_N$ has not been experimentally observed so far.

\section{Summary}

Ultrasound velocity measurements of CaV$_2$O$_4$ have revealed the elastic-mode-dependent emergence of Curie-type softening above and below $T_s$ in the temperature dependence of the shear elastic moduli $C_{44}(T)$ and $C_{55}(T)$ in the orthorhombic PM phase ($T>T_s$), and $C_{66}(T)$ in the monoclinic PM phase ($T_N<T<T_s$). Softening in $C_{44}(T)$ and $C_{55}(T)$ above $T_s$ can be attributed to the presence of onsite Jahn-Teller-type and intersite ferro-type orbital fluctuations, which arise respectively in the inequivalent V2 and V1 sites. Softening in $C_{66}(T)$ below $T_s$ can be attributed to a dimensional spin-state crossover, from quasi-one-dimension to three-dimensions, driven by the spin-lattice coupling along the inter-zigzag-chain orthogonal direction. Further experimental and theoretical studies are indispensable if the orbital-lattice order below $T_s$ and the spin-lattice order below $T_N$ in the unique orbital-spin zigzag chain system of CaV$_2$O$_4$ are to be revealed.

\section{Acknowledgments}

This work was partly supported by Grant-in-Aid for Scientific Research (C) (Grant No. 17K05520) from MEXT of Japan, and by Nihon University College of Science and Technology Grant-in-Aid for Research. The research at Ames Laboratory was supported by the U.S. Department of Energy, Office of Basic Energy Sciences, Division of Materials Sciences and Engineering.  Ames Laboratory is operated for the U.S. Department of Energy by Iowa State University under Contract No. DE-AC02-07CH11358.


\begin{thebibliography}{prb}
\bibitem{Niazi2} According to Ref. [\cite{Niazi}], the relative changes of the orthorhombic axis lengths in the monoclinic crystal phase are at most on the order of $\sim 10^{-5}$ for $\Delta a/a$ and $\Delta b/b$, and $\sim 10^{-4}$ for $\Delta c/c$.
\end{thebibliography}
\end{document}